\documentclass[12pt, a4paper]{article}
\pdfoutput=1
\usepackage[square, comma, sort&compress]{natbib}
\usepackage{amsmath}
\usepackage{graphicx}
\usepackage{enumitem}

\def\lsim{\mathrel{\raise.3ex\hbox{$<$\kern-.75em\lower1ex\hbox{$\sim$}}}}
\def\gsim{\mathrel{\raise.3ex\hbox{$>$\kern-.75em\lower1ex\hbox{$\sim$}}}}

\begin{document}

\thispagestyle{empty}

\sf
\centerline{\Huge Probing leptonic CP phases in LFV processes}

\vspace{7mm}

\centerline{\large Borut Bajc$^1$, Miha Nemev\v sek$^{1,2}$ and Goran Senjanovi\' c$^{3,1}$}

\vspace{1mm}

\centerline{ {\it\small $^1$ J.\ Stefan Institute, 1000 Ljubljana, Slovenia} }
\centerline{ {\it\small $^2$ II. \ Institut f\"ur Theoretische Physik, Universit\"at Hamburg, Luruper Chaussee 149, 22761 Hamburg, Germany} }
\centerline{ {\it\small $^3$ International Centre for Theoretical Physics, 34100 Trieste, Italy} }

\vspace{10mm}

\centerline{\large\sc Abstract}
\begin{quote}
\small

We study a CP and T violating triple (spin) correlation in the muon to electron conversion
in nuclei in the context of the seesaw mechanism. After concluding that the results are negative for all
three seesaw types, we turn to the left-right symmetric theories as the original source of seesaw. We find
that in general this correlation is of order one which offers a hope of observing CP violation in lepton flavor
violating processes for a $L-R$ scale below around $10-30 \text{ TeV}$. We discuss the conditions that could render to (unlikely) conspiracies as to suppress the CP violating effects.

\end{quote}
\rm

\newpage

\setcounter{page}{1}

\section{Introduction}

Probing CP phases is a great challenge of neutrino physics. They can be manifest in CP even processes at colliders 
\cite{Kadastik:2007yd} and in neutrinoless double beta decay  \cite{Strumia:2006db}
or as CP odd in neutrino-antineutrino oscillations \cite{deGouvea:2002gf}.
Another possibility is to study the LFV processes with best experimental limits, $\mu \to e \, \gamma$, $\mu \to 3 \, e$ and $\mu \to e$ conversion in nuclei. These are very rare processes and as such provide an ideal window into physics behind neutrino masses and mixings. While the total decay rates themselves are sensitive to CP phases, additional information can be obtained by studying correlations between the polarization of the initial muon state and the final state particles. These are the so-called triple product correlations, studied at length in the literature as a probe of CP violation \cite{Rindani:1994ad, Yuan:1994fn} and recently revisited in the context of leptonic CP violation \cite{Farzan:2007us, Ayazi:2008gk, Davidson:2008ui, Farzan:2009us}. Particularly important
is $\mu \to e$ conversion, for there is a serious proposal \cite{Ankenbrandt:2006zu,P21Jparc,P20Jparc} 
to improve its sensitivity by four to six orders of magnitude, which would
bring it to an unprecedented precision. This process is thus worth a particular attention from the theoretical point of view and is the focus of our interest.

  We study here the P, CP and T violating triple correlation of muon and electron spins, and the electron momentum in the
  context of the so-called seesaw mechanisms. 
Assuming a single type of mediators, one conventionally speaks of three types of seesaw. Type I  \cite{seesawI}, 
when the mediators are fermionic singlets called right-handed neutrinos, type II   \cite{Magg:1980ut}
when the mediator is an $SU(2)$ triplet scalar particle and type III \cite{Foot:1988aq}
when the fermionic mediators are $SU(2)$ triplets. Strictly speaking, these simple scenarios have no strong theoretical
motivation in themselves.  The types I and II emerge naturally in the context of $L-R$ symmetric theories, such
as Pati - Salam theory \cite{patisalam} or $SO(10)$ grand unified theory, and type III in the context of a minimal realistic $SU(5)$ theory \cite{Bajc:2006ia}.

For that reason, we cover all the three cases. Our findings are negative, unless one is willing to go to a small corner of parameter space. On the other hand, it is much more appealing to have a real theory that connects the smallness of neutrino mass to different physical phenomena. A natural example is provided by the left-right symmetric theories  \cite{leftright},
which historically have led to the seesaw picture for neutrino mass. In contrast to the simple-minded seesaw approach, in this case our findings are rather optimistic, as long as the scale $M_R$ of left-right symmetry breaking (or at least some of its remnants) lies below $10-30 \text{ TeV}$. Of course, if $M_R$ is in the TeV region, this would be quite exciting from the collider prospect point of view.

 

This paper is organized as follows. In the next section, we discuss $\mu \to e $ conversion for the three types of seesaw. In section 3, we repeat the exercise for the left-right symmetric theories, where we also comment on the prospect for the other two important processes, $\mu \to e \, \gamma $ and $\mu \to 3 \, e$.

\section{$\mu \to e$ conversion: leptonic CP phases in the seesaw picture}

$\mu \to e$ conversion in nuclei provides the best experimental limit on lepton flavor violating processes 
\cite{Dohmen:1993mp, Bertl:2006up}

\begin{eqnarray}
B(\mu \text{ Ti}\to e \text{ Ti})&\le&4.3\times 10^{-12},\\
B(\mu \text{ Au}\to e \text{ Au})&\le&7\times 10^{-13},
\end{eqnarray}

\noindent
where

\begin{equation}
B(\mu \text{ N}\to e \text{ N})\equiv\frac{\Gamma(\mu \text{ N}\to e \text{ N})}{\Gamma(\mu \text{ N}\to{\rm capture})}.
\end{equation}
A more stringent bound with Titanium was reported \cite{Wintz:2000qv} $B(\mu \text{ Ti} \to e \text{ Ti}) \leq 6\times 10^{-13}$, but has never been published.

Due to nuclear physics effects the theory of $\mu \to e$ conversion is rich, see for example  \cite{Czarnecki:1997pa,Czarnecki:1998iz,Kuno:1999jp,Kitano:2002mt}. 

A natural quantity that probes CP phases is the P, CP and T violating triple 
correlation of spins and electron momentum:

\begin{equation*}
\left(\vec S_\mu\times\vec S_e\right).\,\vec P_e.
\end{equation*}

To illustrate what happens, let us imagine for the moment that the effective operator, responsible for $\mu\to e$ conversion, takes a simple single Lorentz structure form

\begin{equation}
{\cal L}_{eff}=G_F\sum_{q=u,d}\left(A_L\bar e_L\gamma^\mu\mu_L+A_R\bar e_R\gamma^\mu\mu_R\right)\left(V_L^q\bar q_L\gamma^\mu q_L+V_R^q\bar q_R\gamma^\mu q_R\right)+\text{h.c.}.
\end{equation}

The coefficient of the triple correlation turns out to be proportional to \cite{Ayazi:2008gk}

\begin{equation}
\delta_{CP}=\frac{Im(A_L^*A_R)}{|A_L|^2+|A_R|^2}.
\end{equation}

The result is easily understood on physical grounds.  Since for a single helicity of the electron the spin would be proportional to its motion, the spin of the  electron being perpendicular to its motion in this correlation requires the presence of both, $A_L$ and  $A_R$. CP violation then requires a relative phase between  $A_L$ and $A_R$. The same reasoning applies to the situation when more than one operator is present, as can be seen in \cite{Ayazi:2008gk} and can be (un)easily generalized to an arbitrary case of such operators. Hereafter, we will use the notation $A_L$ and  $A_R$ to denote any operator that involves $e_L$ and $e_R$, respectively. Notice that our notation, consistent with electron (and muon) chirality, is different from \cite{Kitano:2002mt} (and \cite{Ayazi:2008gk}) who use the subscript L for the scalar and vector interactions, but use R for the tensor one for the same L chirality of the electron. 

It is straightforward to see that the seesaw mechanisms lead to a negligible triple correlation.
The crucial point is that the different types of seesaws are characterized by one common aspect: only left-handed charged leptons are involved. These interactions are respectively
\begin{itemize}

\item
%
 $\ell H F_{\text{new}}$,
%

where $F_{\text{new}}$ is a singlet fermion (called right-handed neutrino) in the type I and a $SU(2)$, $Y=0$ fermion triplet in the type III, $\ell$ stands for the usual leptonic doublet and $H$ the standard model Higgs doublet; 

\item
%
 $\ell \ell \Delta$,
%

where $\Delta$ is an $SU(2)$ triplet, $Y=2$ scalar in the case of type II seesaw. 
\end{itemize}

\noindent This simple fact provides the cornerstone for our reasoning in what follows.







\begin{figure}
\begin{center}
	\includegraphics[height=4cm]{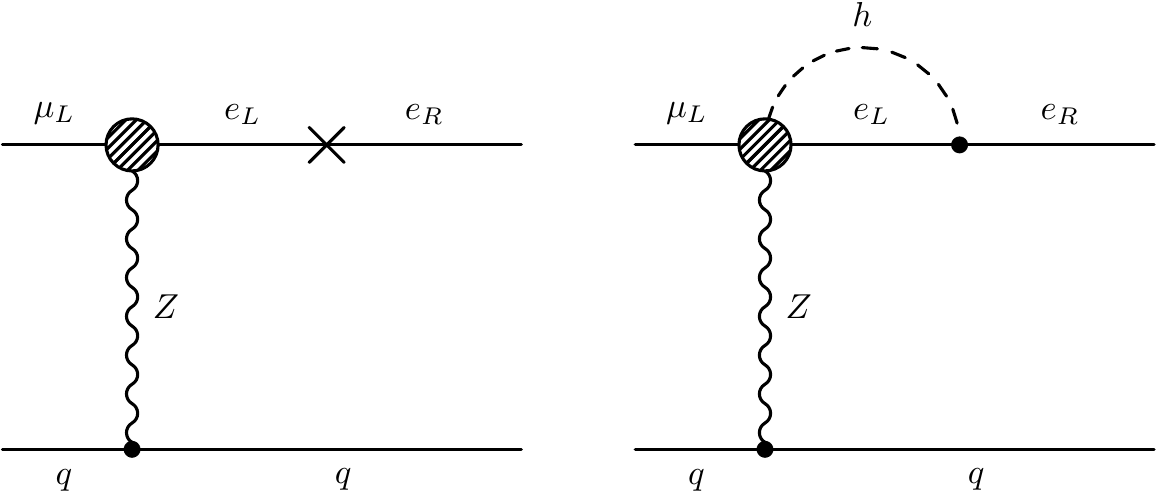}
\end{center}
\caption{ \label{figure:pic1} The right-handed electron contributes to $\mu \to e$ conversion only through the 
mass insertion on the left or through the Higgs loop on the right.}
\end{figure}

 We must bring $A_R$ into the game. The simple mass insertion on the external electron leg does not suffice, for then $A_R$ has the same form as $A_L$. In that case 

\begin{equation}
A_R=\frac{m_e}{m_\mu}A_L
\end{equation}

\noindent
and thus $\delta_{CP}=0$. To obtain a nontrivial imaginary part, one has to bring in the Higgs exchange, which implies $A_R$ being loop suppressed compared to $A_L$. This is illustrated in fig. \ref{figure:pic1} for the case of $Z$ exchange contribution to $\mu \to e$ conversion. In short, it is easy to estimate 

\begin{equation}
\delta_{CP}\approx \frac{\alpha}{\pi}\frac{m_e}{M_W}\approx 10^{-7},
\end{equation}

\noindent
where $m_e/M_W$ is simply due to the electron Yukawa coupling.

Independently of the type of the seesaw, the prospect of measuring CP violation and probing the CP phases is hopeless, even if one were to arrive at $10^{-18}$ upper limit for the branching ratio.

The picture of seesaw is somewhat simple minded and it is instructive to see what happens in a well defined theory. 
We can guess the answer from what we have learned here: if at low energies you are left with only the seesaw, whatever the type(s), the CP violating correlations vanish. An example of such a theory is provided by a minimal extension of the original $SU(5)$ theory that can simultaneously account for the unification of gauge couplings and neutrino mass. It is based on an addition of an adjoint $24_F$ fermionic representation \cite{Bajc:2006ia},
which leads to the hybrid type I and type III seesaw and no other low energy manifestation. It then predicts, as above, no CP violating effects.

In the next section we discuss the left right symmetric theory which originally led to the seesaw mechanism. Here, on
the contrary, you would expect a large contribution to $\delta_{CP}$, for both left and right electrons are present and 
$L-R$ symmetry is broken.

\section{The left-right symmetric model}

We focus here on the minimal left-right symmetric theory with the seesaw mechanism \cite{seesawlr}. This class of models is characterized by both type I and type II seesaw. They are defined by the minimal fermionic assignment 
and the following fields in the Higgs sector:

\begin{equation}
\Phi(2,2,0), \quad \Delta_L(3,1,2), \quad \Delta_R(1,3,2)
\end{equation}

\noindent
under $SU(2)_L\times SU(2)_R \times U(1)_{B-L} $. This allows for new Yukawa couplings of $\Delta$'s with the
leptons

\begin{equation}
{\cal L}_{\Delta} = Y_{\Delta} (\ell_L \ell_L \Delta_L + \ell_R \ell_R \Delta_R) + \text{h.c.}.
\end{equation}

The parity breaking vev

\begin{equation}
\langle\Delta_R\rangle \simeq M_{W_R},  \quad \langle\Delta_L\rangle= 0
\label{mwr}
\end{equation}

\noindent
is responsible for the original breaking down to the SM symmetry,  and the vev of 
the bi-doublet 

\begin{equation}
\langle\Phi\rangle=M_L
\end{equation}

\noindent
completes the symmetry breaking.
This will induce an effective potential for $\Delta_L$, in the symbolic notation

\begin{equation}
V_{\Delta_L} = M_{\Delta_L}^2 \Delta_L^2 + \alpha \Delta_L \Phi^2 \Delta_R + \ldots ,
\label{Vdelta}
\end{equation}
 
 \noindent which leads to a small vev for $\Delta_L$
 
 \begin{equation}
\langle\Delta_L\rangle = \alpha \frac{\langle\Phi\rangle^2  \langle\Delta_R\rangle }{M_{\Delta_L}^2},
\label{deltavev}
\end{equation}

\noindent
which is responsible for the type II contribution to the neutrino mass.

The spontaneous breakdown of parity leads to different masses 

\begin{equation}
M_{\Delta_L} \ne M_{\Delta_R} 
\end{equation}

\noindent with, in general 

\begin{equation}
M_{\Delta_L}, \; M_{\Delta_R}, \; M_{\Delta_L} - M_{\Delta_R} \propto \langle\Delta_R\rangle.
\end{equation}

\noindent From  (\ref{Vdelta}) and  (\ref{deltavev}), one can easily find the mixing between  $\Delta_L$ and $\Delta_R$
to be

\begin{equation}
\theta_{\Delta_L\Delta_R} \simeq \frac{\langle\Delta_L\rangle}{\langle\Delta_R\rangle}.
\end{equation}

Since  $\langle\Delta_L\rangle \lsim \text {GeV}$ and $\langle\Delta_R\rangle \gsim \text{TeV}$  \cite{Beall:1981ze},
this mixing has to be small: $\theta_{\Delta_L\Delta_R} \lsim 10^{-3}$. In reality, the limit is much smaller.
Barring the possibility that
neutrino is light due to fine-tuned cancellations between large type I and type II contributions,  one has

\begin{equation}
\theta_{\Delta_L\Delta_R} \simeq \frac{m_{\nu}}{m_N},
\end{equation}

\noindent where 

\begin{equation}
m_N = Y_{\Delta} \langle\Delta_R\rangle
\label{mn}
\end{equation}
   
  \noindent stands symbolically for the right handed neutrino masses and $m_\nu$ stands for the combined contribution
 of type I and type II neutrino masses. We will see in the next section that in order to have an appreciable amount of
 $\mu \to e$ conversion, one needs $m_N \gsim (1-10) \text{GeV}$. In short, we get a much stronger limit
 
\begin{equation}
\theta_{\Delta_L\Delta_R} \lsim 10^{-9}.
\end{equation}
 
 It is also known that the mixing between the left and right gauge 
 bosons must be small \cite{Langacker:1989xa}
 
 \begin{equation}
\theta_{W_L W_R} \lsim 10^{-2}.
\end{equation}
 
\noindent Thus, in our estimates, we can safely ignore these mixings between left and right sectors of the theory. This substantially simplifies the analysis.

\subsection{$L-R$ symmetry and LFV}

The charged fields in $\Delta_{L,R}$ play an important role in LFV, as we will stress below.

\begin{figure}
\begin{center}
	\includegraphics[height=4cm]{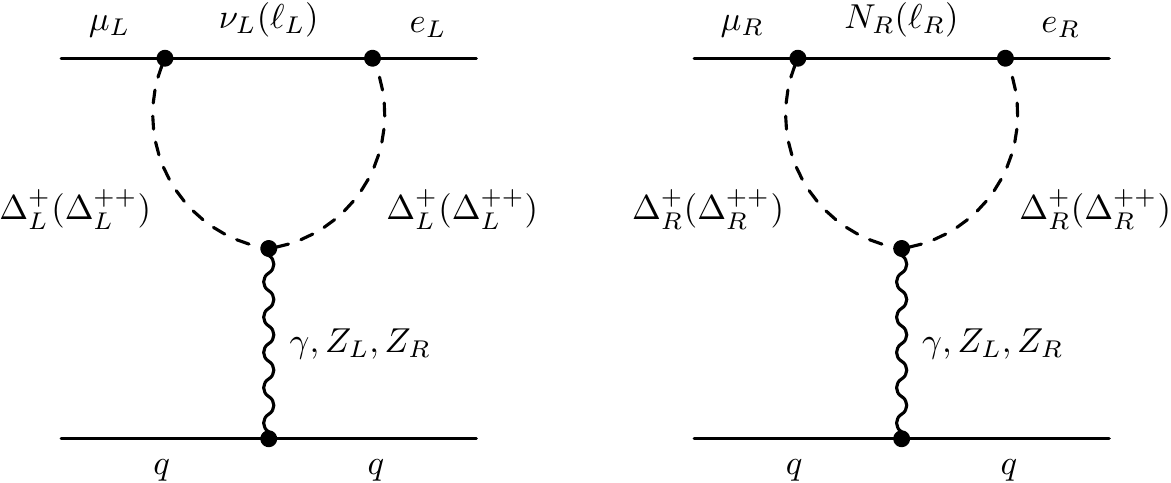}
\end{center}
\caption{ \label{deltalr} The contributions of the scalar triplets $\Delta_L$ and $\Delta_R$ to the typical penguin diagrams for $ \mu \to e$ conversion.}
\end{figure}

At first glance, one could naively fear that $\delta_{CP}\approx(M_L/M_R)^2$, as is typical of processes which need new physics associated with the $M_R$ scale\footnote{$M_R$ denotes generically the scale of $L-R$ symmetry breaking and hereafter stands for the leading contribution(s) 
from either $W_R$ or the $\Delta$ states.}. The point is that in the standard model $A_L$ is negligible due to the GIM mechanism

\begin{equation}\label{gim}
A_L(W_L)\propto\frac{\Delta m_\nu^2}{M_L^2}\leq 10^{-25}.
\end{equation}

This in turn brings the fear that $A_L/A_R$ is vanishingly small which would again lead to a small $\delta_{CP}$. However, $A_L$ has a new source: the $\Delta_L$ exchange (fig. \ref{deltalr}), where both $\Delta_L^+$ and $\Delta_L^{++}$ participate. These new contributions are {\it not} GIM suppressed since the $Y_\Delta$ couplings violate leptonic flavor. 
From the diagram in fig. \ref{deltalr}, we have

\begin{equation}
A_L(\Delta_L)\approx A_R(\Delta_R)\approx \left(\frac{M_L}{M_R}\right)^2\frac{\alpha}{\pi} \, Y_\Delta^2.
\end{equation}

If $A_L$ and $A_R$ have the same complex phase structure, the question is then whether $Im(A_LA_R^*)$ vanishes or not. We discuss this below.

The crucial point is the complete absence of $L-R$ symmetry in the neutrino sector. While the standard model $A_L$ contribution is negligible (\ref{gim}), the $W_R$ exchange (coming from the box diagrams and the 
penguins with the $Z$ and $\gamma$ similar to those in fig. \ref{deltalr})  gives an analog

\begin{equation}
A_L(W_L)\approx 0,\;\;\;\;A_R(W_R)\approx\left(\frac{M_{L}}{M_{R}}\right)^2
\frac{\alpha}{\pi}\left(\frac{m_N}{M_{W_R}}\right)^2.
\end{equation}

\noindent
The $(m_N/M_{W_R})^2$  stands symbolically for a right-handed GIM factor,
which is not a priori small. Barring an accidental cancellation 

\begin{equation}
\delta_{CP}={\cal O}(1).
\end{equation}

This offers a great hope of observing CP violation and probing CP phases for a not too high $M_R$ scale. One can easily estimate (up to the flavor mixings, expected to be large)

\begin{equation}
B(\mu \to e\;\;{\rm conversion})\approx\left(\frac{\alpha}{\pi}\right)^2\left(\frac{M_L}{M_R}\right)^4\left(\frac{m_N}{M_{W_R}}\right)^4.
\end{equation}


In order to be able to study the CP violating correlation, one should have a sufficient number of events and so it is reasonable to
demand a large enough value $B(\mu\to e\;\;{\rm conversion}) \gsim10^{-14}$. From the lower limit 
$M_R \gsim 100 \text{ GeV}$, one gets

\begin{equation}
3\times 10^{-3} \lsim \frac{m_N}{M_{W_R}} \lsim 1,
\end{equation}

\noindent where the upper limit simply means perturbativity  from (\ref{mwr}) and  (\ref{mn}).

From the same equations, one gets roughly $M_R \lsim Y_{\Delta} \, 300 \, M_L$, which implies an absolute
upper limit of $M_R \lsim 10 \text{ TeV}$.
  For a detailed calculation and a comprehensive
study of LFV in $L-R$ symmetry, see \cite{Cirigliano:2004mv}.

Let us illustrate what is going on by discussing the diagram with $\Delta_R^+$ of fig. \ref{deltalr} with $N$ in the loop.
It can be shown to have a flavor structure of the form

\begin{equation}
U_R^{\dagger}m_N^2f(m_N^2/M_R^2)U_R,
\end{equation}
where $U_R$ is the leptonic mixing matrix in the right-handed sector (the analog of the PMNS matrix $U_L$), $m_N$ stands for the diagonal mass matrix of right-handed neutrinos and the function $f(m_N^2/M_R^2)$ denotes the loop dependence. This is immediately obtained from the fact that the (original, non-diagonal) mass matrix of right-handed neutrinos $M_N$ is proportional to $Y_\Delta$. The diagram with $\Delta_L$ of fig. \ref{deltalr} has the flavor structure \cite{marinero}

\begin{equation}
U_L^{\dagger}U_{\nu N}^{\dagger}m_N^2f(0)U_{\nu N}U_L,
\end{equation}
where $U_{\nu N}$ is the mismatch between the unitary matrices that diagonalize $\nu$ and $N$ mass matrices and $f(0)$ indicates we have light neutrinos in the loop. More precisely, 

\begin{equation}
U_{\nu N}=U_\nu^\dagger U_N,
\end{equation}
with

\begin{equation}
U_\nu^TM_\nu U_\nu=m_\nu\;\;\;,\;\;\;U_N^TM_NU_N=m_N.
\end{equation}

It is easy to show that the same structure emerges from a diagram with  $\Delta_R^{++}$ in fig. \ref{deltalr},  except that one
has $f(0)$ in both left and right cases, since the fermions in the loop are light charged leptons.

In general, $U_{\nu N}$ is an arbitrary matrix, since in general both type I and type II seesaws contribute to light neutrino masses. This makes it hard to make any prediction and even to disentangle the phases if the triplet correlation were to be measured. The situation simplifies considerably, if type II dominates, in which case $U_{\nu N}=I$ and this process probes the relative phases in left and right sectors. In order to do this, the masses of right-handed neutrinos will have to be probed by then. This could in principle be achieved in colliders, see below. 

We have established the important result that $L-R$ symmetric theory predicts a sizable amount of CP violation in the triple correlation, even if $W_R$ is out of LHC reach. It could happen though, that this phase vanishes accidentally, but that would involve fine-tuning between large contributions. 

  The question is then under which physical conditions does the $\delta_{CP}$ phase vanish? First, it requires the suppression of $W_R$ loop due to the asymmetry between $\nu$ and $N$, so one possibility is a  large mass for $W_R$, 
 $M_{W_R}\gg  10 \text{ TeV}$. We can call this $W_R$ decoupling. If $M_{\Delta_L}\approx M_{\Delta_R}\approx M_{W_R}$,  we are back to the SM and vanishing $\mu \to e$ conversion, so assume that only $W_R$ is decoupled.

We can have then

\begin{enumerate}[label=a)]
	\item $M_{\Delta_L}\approx M_{\Delta_R}\ll M_{W_R}$. In this case $\delta_{CP}$ would vanish naturally under three conditions:
\end{enumerate}

\begin{enumerate}[label=a\arabic{*})]
	\item same mixings of left and right charged leptons, which is equivalent to the Hermitean mass matrix of charged leptons. This is known as manifest $L-R$ symmetry.  While this may happen in the minimal $L-R$ model, it is by no means generic.  
	\item same couplings in left and right neutrino sectors, which is automatic in type II seesaw.
	\item right handed neutrinos much lighter than $\Delta_R^+$. If all three conditions are satisfied, $U_L=U_R$ and $\delta_{CP}\to 0$. 
\end{enumerate}

\begin{enumerate}[label=b)]
	\item $M_{\Delta_L}\ll M_{\Delta_R}$ (or vice versa), then $A_R/A_L\to 0$ ($A_L/A_R\to 0$) 
and again $\delta_{CP}\to 0$.
\end{enumerate}
While such conspiracies are possible, they are quite unlikely. 

What about other LFV processes such as $\mu\to e\gamma$ and $\mu\to ee\bar e$? A quick glance assures one of the completely analogous situation to $\mu\to e$ conversion: in general, one expects large CP violating effects here, too. Simply, the CP asymmetries of the kind discussed here involve similarly left and right amplitudes that are comparable as we have argued above. Of course,
what is needed is a serious improvement in experimental branching ratios for these processes in order to have enough events to probe the polarization of the outgoing electron. 

\subsection{Other manifestations of low $M_R$}

\begin{enumerate}[label=\roman{*})]

\item{Colliders}

A light $W_R$ would have striking signatures at colliders such as the LHC. Through the production of $N$ one gets same 
sign di-leptons \cite{Keung:1983uu} 
as an indication of lepton number violation (LNV) and could observe directly both parity restoration and the origin of neutrino mass. Through LFV channels one could also probe the CP phases in this situation. This of course requires $M_{W_R}\lsim 3-4$ TeV  
\cite{Ferrari:2000sp}. There have been recent claims of $M_{W_R}\gsim 4$ TeV \cite{Zhang:2007da} (or even $M_{W_R}\gsim 10$ TeV \cite{Xu:2009nt})
 in the minimal theory, but these limits depend on the definition of $L-R$ asymmetry and its manifestness.
Recall that $L-R$ symmetry can be P, as in the original works, or C as it happens in $SO(10)$. The authors of  \cite{Zhang:2007da, Xu:2009nt}
use P, but one must check for C, too. Furthermore, they argue in favor of almost manifest $L-R$ symmetry, which is open to questioning.

In the case of the $W_R$ decoupling,  the lighter $\Delta$ states could be observed at the LHC, and the doubly charged 
fields would have spectacular signatures of the pairs of same charge leptons and anti-leptons. Again, LHC and LFV
could provide complementary information.

\item {Neutrino-less double beta decay}

A light $W_R$, with a mass of a few TeV, and $m_N$ between 100 GeV - TeV can easily dominate $\beta\beta 0\nu$ \cite{Mohapatra:1980yp}. Of course, its contribution depends on the right-handed leptonic gauge mixing matrix $U_R$
and so $\beta\beta 0\nu$ provides another source of information on the phases in $U_R$.
\end{enumerate}

\section{Summary and outlook}

   Measuring CP violation in the leptonic sector is a great challenge, since one needs to probe rather feeble effects.
Rare LFV decays may be important in this regard, especially $\mu \to e$ conversion in nuclei that could reach unprecedented 
  precision. 
 At the same time,  probing the origin of neutrino mass is as much of a challenge. The dominant belief is that small neutrino
 masses stem from the seesaw mechanism, which in the minimal scenario, comes in three different types.  Motivated by this,
 we have studied here CP violating triple correlation in $\mu \to e$ conversion for all the three seesaw types. Our findings are 
 rather negative, for the relevant CP  phase ends up being extremely tiny $\delta_{CP} \leq 10^{-7}$.
 
 On the other hand, the simple seesaw mechanism, where one adds just a particle(s) to the SM only to get massive
 neutrinos, is not very convincing, especially since the properties of the new particles are completely arbitrary. It is
 more appealing to have a theory behind, and an original example is provided by $L-R$ symmetric theories, where small
 neutrino mass is tied to (almost) maximal parity violation in the SM.  These models not just give possibly large LFV
 effects, but also naturally predict a large CP phase, of order one, as long as the $M_R$ scale is not terribly large, roughly below $10-30 \text{ TeV}$. This is welcome news: CP violation in LFV can shed light on the theory behind neutrino masses.
 Observable effects would clearly discredit a simple seesaw scenario, whatever the type and would help probe
 the nature of seesaw in the minimal  $L-R$ symmetric model. 
 
  All of this holds true under one important caveat, i.e. that the decaying muons are completely polarized.  However, even if negative muons in the beam are 100 \% polarized, they are depolarized during their atomic cascades down to the 1 s ground state.
  There is a small residual polarization, of about 15 \% \cite{evseev} in nuclei with zero spin, but is much smaller when nuclei carry spin. 
  They must be re-polarized, and one way is to have a polarized nucleus target \cite{Nagamine:1974yu, Kuno:1987dp}.
  Clearly, the CP violating triple correlation, which we estimate to be of order one in $L-R$ theory, must be weighed by the amount of the actual muon polarization.
  
   Suppose that the large CP violating effect is observed in future in $\mu\to e$ conversion. What will be the next step before
   claiming that one is seeing the $L-R$ model and not, say the low energy supersymmetry that its devotees will claim? Surely one CP phase is not enough. One must compute both CP conserving and
   violating rates for different nuclei in order to determine the type of operator(s) which is (are) responsible for this process 
   \cite{Cirigliano:2009bz}.  This can in turn help determine which particle(s) mediates the conversion. It would help 
   if LHC were to discover $L-R$ symmetry and measure the masses and mixings of the new states in the theory (or even
   better if we were to hit a jackpot).  Needless to say, one should compute the rates and CP correlations for other LFV rare 
   decays, such as $\mu\to e \gamma$ and $\mu\to 3 e$. We plan to address this in near future.


\section*{Acknowledgments}

We would like to thank Tsedenbaljir Enkhbat for useful discussions and Yasuhiro Okada for correspondence on the topic of muon to electron conversion. The work of M.N. and B.B. was supported by the Slovenian Research Agency and the work of M.N. also by the Deutsche Forschungsgemeinschaft via the Junior Research Group ÒSUSY PhenomenologyÓ within the Collaborative Research Centre 676 ÒParticles, Strings and the Early UniverseÓ. The work of G.S. was partially supported by the EU FP6 Marie Curie Research and Training Network ÓUniverseNetÓ (MRTN-CT-2006-035863) and by the Senior Scientist Award from the Ministry of Science and Technology of Slovenia.

\end{document}